\documentclass[aps]{revtex4-2}

\usepackage{eurosym}
\usepackage{graphicx}
\usepackage[colorlinks=true, pdfstartview=FitV, linkcolor=blue, citecolor=red, urlcolor=magenta]{hyperref}
\usepackage{graphicx}
\usepackage{latexsym}
\usepackage{amsmath}
\usepackage{amsfonts}
\usepackage{amssymb}
\usepackage{verbatim}
\usepackage{braket}
\usepackage{extarrows}
\usepackage{hyperref}
\usepackage{orcidlink}
\usepackage{geometry}
\usepackage[english]{babel} 
\usepackage{tikz}
\usetikzlibrary{arrows.meta, decorations.pathmorphing, 3d,calc}

\begin{document}

\title{Casimir effect for a massive scalar field confined between parallel plates with a spatially varying effective mass}

\author{R. L. Araújo Xavier}
\email{rlax2@academico.ufpb.br}

\author{M. H. B. Chaves}
\email{mhbc@academico.ufpb.br}

\author{E. R. Bezerra de Mello~\orcidlink{0000-0002-6115-5052}}
\email{emello@fisica.ufpb.br}

\author{Herondy~Mota~\orcidlink{0000-0002-7470-1550}}
\email{hmota@fisica.ufpb.br}

\affiliation{Departamento de F\'{\i}sica, Universidade Federal da Para\'{\i}ba,\\
Caixa Postal 5008, Jo\~{a}o Pessoa, Para\'{\i}ba, Brazil}

\begin{abstract}
We investigate the Casimir effect for a massive real scalar field confined between two perfectly reflecting parallel plates in the presence of a position-dependent effective mass, a mechanism for coupling a scalar background to a scalar field. Exact normal modes are obtained by solving the corresponding Klein-Gordon equation, leading to a transverse energy spectrum that exhibits a characteristic Landau-like structure despite the absence of an external magnetic field. Upon quantization of the field, the vacuum energy is evaluated by means of generalized zeta-function regularization together with an appropriate renormalization procedure. The renormalized vacuum energy naturally separates into a Landau-like contribution and an additional term induced by the spatial dependence of the effective mass. We show analytically and numerically that both contributions are exponentially suppressed in the strong-coupling regime. In the opposite limit, the Landau-like contribution smoothly reproduces the standard vacuum energy for a massive scalar field confined between parallel plates, whereas the additional contribution becomes singular owing to the restricted domain of validity of the exact spectrum. Except in the vicinity of this singular limit, the vacuum energy is shown to be dominated by the Landau-like sector. Our results establish a direct connection between position-dependent effective masses and boundary-induced quantum vacuum phenomena, providing a new exactly solvable framework for the investigation of Casimir effects in spatially inhomogeneous relativistic systems.
\end{abstract}

\maketitle

\section{Introduction}
\label{intro}
The Casimir effect is one of the most remarkable manifestations of quantum vacuum fluctuations induced by the imposition of boundary conditions on relativistic fields. It was first predicted by Casimir in 1948~\cite{casimir1948attraction} for the electromagnetic field confined between two perfectly conducting parallel plates, leading to the existence of an attractive force between them. The first experimental attempt to detect this phenomenon was carried out by Sparnaay in 1958~\cite{Sparnaay1958}. Although his measurements were consistent with the existence of the Casimir force, the experimental uncertainties were too large to provide a quantitative confirmation. A precise verification of the theoretical prediction became possible only several decades later through the pioneering experiment of Lamoreaux in 1997~\cite{lamoreaux1997demonstration,lamoreaux1998erratum}, followed by a series of high-precision measurements performed by Mohideen and collaborators~\cite{Mohideen:1998iz,PhysRevLett.82.4380}. These experiments, based on the sphere-plane configuration, established the Casimir effect as one of the most important observable consequences of quantum vacuum fluctuations. Subsequently, the first direct observation of the Casimir force between two parallel conducting plates was achieved by Bressi \textit{et al.}~\cite{bressi2001experimental, bressi2002measurement}, providing an important confirmation of Casimir's original proposal.

Although originally formulated for the electromagnetic field, the Casimir effect has since been investigated in a broad variety of quantum field-theoretical settings. In particular, scalar fields have received considerable attention because they provide a comparatively simple framework for studying vacuum fluctuations and boundary-induced phenomena while preserving the essential physical features of the Casimir effect. Besides their theoretical relevance, scalar field models have applications in several branches of modern physics, including cosmology, condensed matter, and effective quantum field theories~\cite{bordag2009advances,milton2001casimir,mostepanenko1997casimir}. Nevertheless, most investigations have focused on the influence of boundary conditions and external electromagnetic backgrounds. Comparatively less attention has been devoted to the role played by external scalar backgrounds, despite the fact that scalar interactions are commom in modern theoretical physics, ranging from relativistic quantum mechanics to effective field theories.

Unlike electromagnetic interactions, there is no unique prescription for incorporating external scalar interactions into relativistic wave equations. Depending on the physical problem under consideration, different realizations have been proposed in the literature. One of the earliest approaches was introduced by Dosch \textit{et al.}~\cite{Dosch1971} and subsequently developed by Soff \textit{et al.}~\cite{Soff1973}, in which a non-electromagnetic scalar interaction is incorporated through the replacement $m\rightarrow m+S(x)$, allowing the scalar background to modify the mass term of the relativistic wave equation. Alternative formulations have also been investigated in different contexts, as discussed in Refs.~\cite{Bordag1996, CavalcantideOliveira:2006row, Bakke:2014qux, Vitoria:2015vee, vitoria2016relativistic, Vitoria:2016bwq}. These examples illustrate that the implementation of scalar backgrounds is not unique and depends on the underlying physical assumptions.

Motivated by this scenario, in the present work we adopt a different perspective by describing the external scalar interaction through a spatially varying effective mass. This approach is motivated by the fact that position-dependent effective masses constitute a well-established concept in several areas of physics. In semiconductor heterostructures, for example, the effective mass of charge carriers varies across material interfaces owing to changes in the electronic band structure~\cite{vonRoos:1983zz,PhysRevA.52.1845,Bastard1988}. Similar concepts arise in relativistic mean-field descriptions of nuclear systems, where interactions with background mesonic fields modify the nucleon effective mass~\cite{Serot:1997xg,Serot:1984ey}, as well as in optical and metamaterial analogues, where spatial inhomogeneities can be described through effective medium parameters~\cite{PhysRevB.86.161104}. Inspired by these developments, we consider the simple quadratic realization $m^2\rightarrow m^2+\alpha^2\rho^2$, where $\alpha$ is an effective coupling parameter of the model and $\rho$ is the polar coordinate, defined later in the text. This modification represents an external scalar background encoded in the mass term. A similar idea has also been discussed in the context of relativistic wave equations in Ref.~\cite{Greiner2000}. From a theoretical point of view, a spatially varying effective mass provides one of the simplest phenomenological ways of incorporating spatial inhomogeneities into relativistic wave equations while preserving their local structure.

Besides its mathematical simplicity, this choice possesses an appealing physical feature. In the nonrelativistic limit, the present relativistic model naturally reduces to a harmonic confining potential. Consequently, the parameter $\alpha$ controls the strength of the effective confinement while preserving the relativistic character of the underlying theory. Our main objective is therefore to investigate how such a spatially varying effective mass modifies the spectrum of vacuum fluctuations and, consequently, the vacuum energy for a massive scalar field confined between two parallel plates.

An additional motivation for the present model stems from the structure of its normal-mode spectrum. As will be shown, the expression for the vacuum energy resembles the relativistic Landau spectrum of a charged scalar particle in a uniform magnetic field, with the parameter $\alpha$ playing a role analogous to the magnetic scale $eB$. An additional contribution, however, distinguishes the present spectrum from the standard Landau levels. Thus, although the underlying physical mechanism is entirely different from that of a magnetic field, the resulting spectrum exhibits a characteristic Landau-like pattern.

This paper is organized as follows. In Sec.~\ref{sec2}, we introduce the model with a position-dependent effective mass, solve the Klein-Gordon equation, obtain its complete solution, and derive the corresponding energy spectrum. We also discuss the nonrelativistic limit of the Klein--Gordon equation. In Sec.~\ref{sec3}, we quantize the real scalar field and calculate the vacuum energy, together with its asymptotic limits. Finally, in Sec.~\ref{sec4}, we present our conclusions. Throughout this paper, we employ natural units in which the reduced Planck constant and the speed of light are set equal to unity, $\hbar=c=1$.

\section{The Klein-Gordon Equation, Eigenfunctions and Energy Spectrum}
\label{sec2}
In this section, we introduce the theoretical model under investigation. It consists of a massive real scalar field endowed with a quadratic position-dependent effective mass. This effective description is implemented through a spatially varying mass term in the Klein-Gordon equation. The corresponding action is given by
\begin{equation}
    S = \frac{1}{2}\int d^{4}x \sqrt{|g|}\,\big[g^{\mu\nu}\partial_\mu\Phi \partial_\nu \Phi - m^2_{\rm eff} \Phi^2\big],
    \label{action}
\end{equation}
where the scalar field is defined over the spacetime coordinates
$x=(t,\rho,\phi,z)$, namely,
$\Phi\equiv\Phi(x)$.
The position-dependent effective mass entering the action is taken as
\begin{equation}
    m_{\rm eff}^{2}=m^{2}+\alpha^{2}\rho^{2},
    \label{eff_mass}
\end{equation}
where $m$ denotes the mass of the scalar field, while $\alpha$ is a positive coupling constant that characterizes the strength of the spatial variation of the effective mass and has mass dimension two. As already discussed, there is no unique prescription for incorporating external scalar interactions into relativistic wave equations. In the present work, we adopt an effective description based on a spatially varying effective mass~\cite{Greiner2000}. This approach is motivated by the widespread use of position-dependent effective masses in several areas of physics, particularly in semiconductor heterostructures, relativistic mean-field descriptions of nuclear systems, and other effective theories describing spatially inhomogeneous media~\cite{vonRoos:1983zz,PhysRevA.52.1845,Bastard1988,Serot:1997xg,Serot:1984ey,PhysRevB.86.161104}.

Throughout this work, the model is formulated in Minkowski spacetime expressed in cylindrical coordinates,
\begin{equation}
    ds^{2}=dt^{2}-d\rho^{2}-\rho^{2}d\phi^{2}-dz^{2},
    \label{MET}
\end{equation}
where $-\infty<t,z<+\infty$, $\rho\geq0$, and $0\leq\phi\leq2\pi$.

Upon varying the action~\eqref{action}, the corresponding equation of motion is obtained as
\begin{equation}
    \left(
    \frac{1}{\sqrt{|g|}}
    \partial_\mu
    \left(
    \sqrt{|g|}
    g^{\mu\nu}
    \partial_\nu
    \right)
    +m_{\rm eff}^{2}
    \right)
    \Phi(x)=0.
    \label{mov.eq.}
\end{equation}
Since $\sqrt{|g|}=\rho$ and considering the metric coefficients in the line element~\eqref{MET}, the above equation reduces to
\begin{equation}
    \left[
    \frac{\partial^{2}}{\partial t^{2}}
    -\frac{1}{\rho}\frac{\partial}{\partial\rho}
    \left(
    \rho\frac{\partial}{\partial\rho}
    \right)
    -\frac{1}{\rho^{2}}
    \frac{\partial^{2}}{\partial\phi^{2}}
    -\frac{\partial^{2}}{\partial z^{2}}
    +m_{\rm eff}^{2}
    \right]
    \Phi(x)=0.
    \label{KGCompleta}
\end{equation}
In order to solve Eq.~\eqref{KGCompleta}, we employ the separable ansatz
\begin{equation}
    \Phi(x)=
    N
    R(\rho)
    e^{-i\omega t+ikz+il\phi},
    \label{ansatz}
\end{equation}
where $N$ is a normalization constant, $k$ is the momentum along the $z$-direction, $l=0,\pm1,\pm2,\ldots$ is the azimuthal quantum number, $\omega$ denotes the eigenfrequency, and $R(\rho)$ is the radial function to be determined. Since our primary interest is the determination of the spectrum, the explicit form of the normalization constant is not required in the following analysis.

After substituting the ansatz~\eqref{ansatz} into Eq.~\eqref{KGCompleta}, we obtain a radial differential equation for $R(\rho)$. An asymptotic analysis of this equation in the limits of small and large radial distances suggests the following form for the radial function:
\begin{equation}
    R(\rho)=\rho^{|l|}e^{-s}F(s),\qquad{\rm with}\qquad s=\frac{\alpha\rho^2}{2}.
    \label{radia_ansatz}
\end{equation}
Substituting the above expression into the radial equation yields
\begin{equation}
    s\frac{d^2 F(s)}{ds^2} + (|l|+1-s)\frac{d F(s)}{ds} - \left( \frac{|l|+1}{2} - \frac{\lambda}{2\alpha}\right)F(s) = 0,
    \label{KDE}
\end{equation}
where
\begin{equation}
    \lambda = \omega^2 - m^2 - k^2.
\end{equation}
Equation~\eqref{KDE} is the Kummer differential equation, whose solution regular at the origin ($\rho=0$) is given by the confluent hypergeometric function~\cite{abramowitz1965handbook}
\begin{equation}
    F = {}_1F_1\left(-n;|l|+1;\frac{\alpha}{2}\rho^2\right),\qquad n=0,1,2,\ldots\,,
    \label{HF}
\end{equation}
where regularity and normalizability require the first argument of the confluent hypergeometric function to be a non-positive integer. Consequently,
\begin{equation}
    n=\frac{\lambda}{2\alpha}-\frac{|l|+1}{2}
    \qquad\Longrightarrow\qquad
    \omega=\sqrt{k^2+m^2+\alpha\big(2n+|l|+1\big)},
    \label{integer_energy}
\end{equation}
which determines the energy spectrum of the system. The complete (unnormalized) solution is obtained by combining Eqs.~\eqref{ansatz}, \eqref{radia_ansatz}, and~\eqref{HF}. It is worth emphasizing that the quantization condition above is well defined only for $\alpha>0$. Thus, the limit $\alpha\to0$ cannot be obtained directly from this solution and must be treated separately.

Therefore, introducing the scalar interaction through the position-dependent effective mass given by Eq.~\eqref{eff_mass} discretizes the transverse spectrum of the scalar field, leading to the quantization of the transverse momenta. In the next section, where the real scalar field is quantized and the corresponding vacuum energy is evaluated, we will show that the resulting expression can be written in a form closely analogous to that encountered in the relativistic Landau problem. In this analogy, the parameter $\alpha$ plays a role analogous to the magnetic scale $eB$, while an additional contribution naturally emerges from the position dependence of the effective mass.

Before concluding this section, let us briefly discuss an interesting feature of the present model. In the nonrelativistic limit, the Klein-Gordon equation~\eqref{mov.eq.}, together with the effective mass~\eqref{eff_mass}, naturally reduces to the Schrödinger equation for a particle subjected to a harmonic confining potential. To demonstrate this, let us restore the constants $\hbar$ and $c$, so that Eq.~\eqref{mov.eq.} assumes the form
\begin{equation}
    \left(-\hbar^2c^2\nabla^2 + m^2c^4+\alpha^2\rho^2\right)\Phi=E^2\Phi,
    \label{KG_NRL}
\end{equation}
where $\nabla^2$ denotes the Laplacian operator in cylindrical coordinates. Introducing the nonrelativistic expansion
\begin{equation}
    E=mc^2+\epsilon,\qquad \epsilon\ll mc^2,
\end{equation}
one readily obtains the Schrödinger equation
\begin{equation}
    \left(-\frac{\hbar^2}{2m}\nabla^2 + V_{\rm H}\right)\Phi=\epsilon\Phi,
    \label{SCH_EQ}
\end{equation}
where
\begin{equation}
    V_{\rm H}=\frac{m\omega_{\rm H}^2\rho^2}{2},
    \qquad {\rm with}\qquad
    \omega_{\rm H}=\frac{\alpha}{mc}.
    \label{H_P}
\end{equation}
Thereby, the quadratic position-dependent effective mass introduced in the relativistic theory naturally gives rise, in the nonrelativistic limit, to an harmonic confining potential whose frequency is completely determined by the coupling parameter $\alpha$. Consequently, the present model possesses two complementary physical interpretations: at the relativistic level, it gives rise to a Landau-like pattern for the energy spectrum of the vacuum, as we shall see in the next section, whereas in the nonrelativistic regime it reproduces the dynamics of a harmonic oscillator. In this limit, the harmonic potential emerges as a relativistic correction, since the potential $V_{\rm H}$ depends explicitly on the speed of light, $c$.
\section{Field Quantization and Vacuum Energy}
\label{sec3}
\begin{figure}[!ht]
    \centering
    \includegraphics[width=0.5\textwidth]{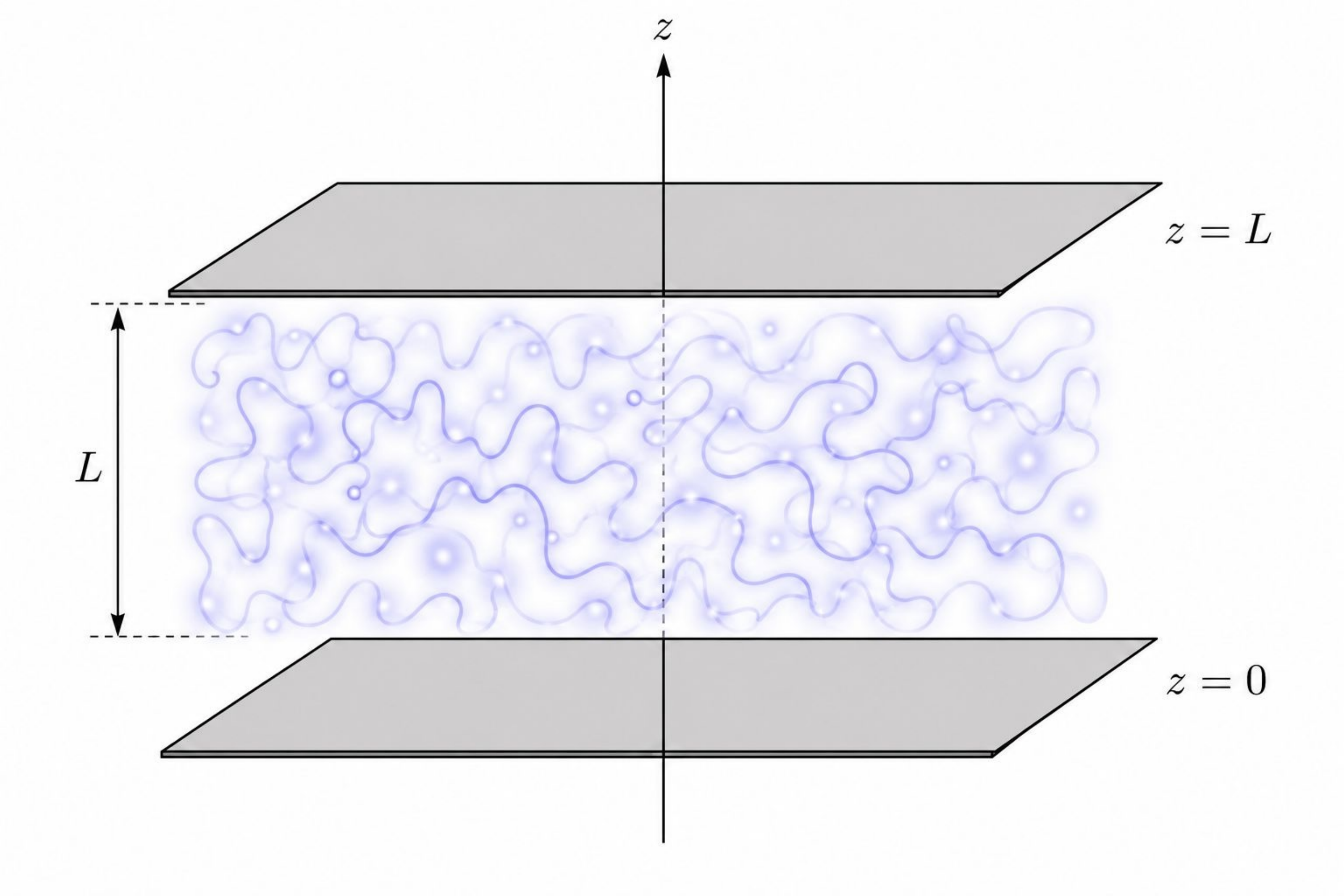}
    \caption{Schematic illustration of two perfectly reflecting parallel plates of area $A$, separated by a distance $L$.}
    \label{figure1}
\end{figure}

We now turn to the calculation of the vacuum energy arising from the modification of the quantum vacuum fluctuations of the real scalar field subject to Dirichlet boundary conditions imposed by two perfectly reflecting parallel plates located at $z=0$ and $z=L$, as illustrated in Fig.~\ref{figure1}. The plates are therefore separated by a distance $L$.

From Eq.~\eqref{ansatz}, we take $\varphi(z)=e^{ikz}$ as the $z$-dependent part of the field. Imposing the Dirichlet boundary conditions
\begin{equation}
    \varphi(z=0)=\varphi(z=L)=0,
    \label{DBC}
\end{equation}
the momentum along the $z$-direction becomes discretized according to
$k\rightarrow k_j$, with $j=1,2,...\,$, and the solution becomes $\varphi(z)=\sin(k_j z)$. Consequently, the energy spectrum in
Eq.~\eqref{integer_energy} takes the form
\begin{equation}
    \omega_{\sigma}=\sqrt{k_j^2+m^2+\alpha\big(2n+|l|+1\big)}\qquad {\rm with}\qquad k_j=\frac{j\pi}{L},
    \label{energyS}
\end{equation}
where $\sigma=(j,n,l)$ denotes the set of quantum numbers labeling the field modes. 
Hence, the eigenfunctions~\eqref{ansatz} take the form
\begin{equation}
    \Phi_{\sigma}(x)=
    N\rho^{|l|}e^{-\frac{\alpha\rho^2}{2}}{}_1F_1\left(-n;|l|+1;\frac{\alpha}{2}\rho^2\right)
    \sin(k_jz)e^{-i\omega t+il\phi}.
    \label{generalS}
\end{equation}

In the quantization procedure, we promote the real scalar field to a quantum field operator $\hat{\Phi}$ and expand it in terms of the positive- and negative-frequency classical solutions given by Eq.~\eqref{generalS}. The coefficients of this mode expansion are promoted to the annihilation, $\hat{a}_{\sigma}$, and creation, $\hat{a}_{\sigma}^{\dagger}$, operators, yielding
\begin{equation}
 \hat{\Phi}(x)=\sum_{\sigma}
 \left[
 \hat{a}_{\sigma}\Phi_{\sigma}(x)
 +
 \hat{a}_{\sigma}^{\dagger}\Phi_{\sigma}^{*}(x)
 \right],
 \label{F_Operator}
\end{equation}
where the summation symbol is defined by
\begin{equation}
 \sum_{\sigma}
 =
 \sum_{j=1}^{\infty}
 \sum_{n=0}^{\infty}
 \sum_{l=-\infty}^{\infty}.
 \label{sum_si}
\end{equation}

The creation and annihilation operators satisfy the canonical commutation relations
\begin{eqnarray}
\big[\hat{a}_{\sigma},\hat{a}_{\sigma'}^{\dagger}\big]
&=&
\delta_{\sigma,\sigma'},\nonumber\\
\big[\hat{a}_{\sigma},\hat{a}_{\sigma'}\big]
&=&
\big[\hat{a}_{\sigma}^{\dagger},\hat{a}_{\sigma'}^{\dagger}\big]
=0,
\end{eqnarray}
where $\delta_{\sigma,\sigma'}$ denotes the Kronecker delta over the set of quantum numbers $\sigma=(j,n,l)$. Consequently, the vacuum state of the scalar field is defined by
\begin{equation}
\hat{a}_{\sigma}|0\rangle=0,
\qquad
\forall\,\sigma .
\end{equation}

Moreover, the Hamiltonian operator associated with the quantized real scalar field takes the known form~\cite{bordag2009advances, milton2001casimir}
\begin{equation}
\hat{H}
=
\sum_{\sigma}
\omega_{\sigma}
\left(
\hat{a}_{\sigma}^{\dagger}\hat{a}_{\sigma}
+\frac{1}{2}
\right),
\label{Hamiltonian}
\end{equation}
where $\omega_{\sigma}$ is given by Eq.~\eqref{energyS}. Taking the vacuum expectation value of the Hamiltonian operator, we obtain the vacuum energy associated with the quantized scalar field,
\begin{eqnarray}
E_{0}
&=&
\langle0|\hat{H}|0\rangle\nonumber\\
&=&
\frac{1}{2}
\sum_{\sigma}
\omega_{\sigma},
\label{VacuumEnergy_D}
\end{eqnarray}
where the summation extends over all allowed field modes defined in Eq.~\eqref{sum_si}.

\subsection{Vacuum energy per unit area of the plates}
\label{sub3A}
Upon considering Eqs.~\eqref{VacuumEnergy_D}, \eqref{sum_si}, and~\eqref{energyS}, we are now in a position to calculate the vacuum energy associated with the system under investigation. Before doing so, however, it is important to note that, when passing from the continuum to the discrete spectrum, an additional factor involving $\alpha$ must be taken into account. Specifically,
\begin{equation}
\frac{V}{(2\pi)^3}\int d^3k
\;\;\rightarrow\;\;
\frac{A\alpha}{2\pi}
\sum_{j=1}^{\infty}
\sum_{n=0}^{\infty}
\sum_{l=-\infty}^{\infty},
\end{equation}
where $V$ denotes the volume of the system and $A$ is the area of each plate.

Therefore, using Eqs.~\eqref{sum_si} and~\eqref{energyS}, the vacuum energy in Eq.~\eqref{VacuumEnergy_D} can be written as
\begin{eqnarray}
E_{0}
&=&
\frac{A\alpha}{4\pi}
 \sum_{j=1}^{\infty}
 \sum_{n=0}^{\infty}
 \sum_{l=-\infty}^{\infty}
\sqrt{k_j^2+m^2+\alpha\big(2n+|l|+1\big)}.
\label{VacuumEnergy1}
\end{eqnarray}

Furthermore, the dependence on $|l|$ can be removed by separating the summation over $l$ into the $l=0$ and $l\geq 1$ contributions. Consequently, one obtains
\begin{eqnarray}
\frac{E_{0}}{A} &=&\frac{\alpha}{4\pi} \sum_{j=1}^{\infty}\sum_{n=0}^{\infty} \sqrt{k_j^2 + m^2 + \alpha\big(2n+1\big)} \nonumber\\
&+& \frac{\alpha}{2\pi} \sum_{j=1}^{\infty}\sum_{n=1}^{\infty}\sum_{l=1}^{\infty} \sqrt{k_j^2 + m^2 + \alpha\big(2n+1 + l\big)}.
\label{VacuumEnergy2}
\end{eqnarray}
The first term exhibits a Landau-like level structure, differing from the conventional Landau spectrum only by an overall factor of two associated with the two degrees of freedom of a complex scalar field, whereas the second term represents an additional contribution. The interpretation of the first contribution becomes particularly transparent upon identifying the parameter $\alpha$ with the magnetic field strength $eB$. Consequently, the spectrum underlying the vacuum energy displays a characteristic Landau-like structure.

Both sums appearing in Eq.~\eqref{VacuumEnergy2} are clearly divergent. Nevertheless, owing to the presence of the parallel plates and the position-dependent effective mass, we expect that a finite physical contribution can be extracted from these divergent expressions. To achieve this, a suitable regularization and renormalization procedure must be employed. One particularly powerful approach is the generalized zeta-function method, which is based on the eigenvalues, $\lambda$, of the differential operator associated with the field equation, such as the one introduced in Eq.~\eqref{mov.eq.}. The generalized zeta function is defined by
\begin{equation}
\zeta(s)=\sum_{\sigma}\lambda_{\sigma}^{-s},
\label{gen_zeta}
\end{equation}
where the summation runs over the complete set of quantum numbers $\sigma$, which may, in general, be either discrete or continuous. The domain of convergence of the series depends on the differential operator and on the dimension of the problem. For a second-order elliptic operator in four-dimensional space, however, the series converges for $\mathrm{Re}(s)>2$ and is regular at $s=0$, admitting an analytic continuation to the remaining values of $s$~\cite{elizalde1995zeta, elizalde1995ten}.

In the present case, the eigenvalues of the differential operator entering the generalized zeta function are identified with the square of the eigenfrequencies given in Eq.~\eqref{energyS}, where $\sigma=(j,n,l)$. After rewriting the vacuum energy in the form of Eq.~\eqref{VacuumEnergy2}, a generalized zeta function can be associated with each of its two contributions. Accordingly, Eq.~\eqref{VacuumEnergy2} can be written as
\begin{eqnarray}
E_{0}(s) &=&\frac{A\alpha}{4\pi} \sum_{j=1}^{\infty}\sum_{n=0}^{\infty} \big(k_j^2 + m^2 + \alpha\big(2n+1)\big)^{-s} \nonumber\\
&+& \frac{A\alpha}{2\pi} \sum_{j=1}^{\infty}\sum_{n=1}^{\infty}\sum_{l=1}^{\infty} \big(k_j^2 + m^2 + \alpha\big(2n+1 + l\big)\big)^{-s}.
\label{VacuumEnergy3}
\end{eqnarray}
The parameter $s$ acts as a regulator, rendering the divergent sums in Eq.~\eqref{VacuumEnergy2} well defined. After evaluating the sums, the divergent contributions can be identified and removed through an appropriate renormalization procedure. Finally, the analytic continuation of the resulting expression is taken to $s=-1/2$, yielding the finite vacuum energy.

The expressions in Eq.~\eqref{VacuumEnergy3} can be further developed by making use of the integral representation
\begin{equation}
   \lambda_{\sigma}^{-s} = \frac{2}{\Gamma(s)} \int_{0}^{\infty} d\tau \, \tau^{2s-1} e^{-\lambda_{\sigma} \tau^2},
   \label{iden}
\end{equation}
where $\Gamma(s)$ is the gamma function~\cite{abramowitz1965handbook}. For convenience, we denote the first term in Eq.~\eqref{VacuumEnergy3}, which exhibits a Landau-like structure, by $E_{\rm L}(s)$, while the second term, representing the additional contribution, is denoted by $E_{\rm c}(s)$.

To proceed, we employ the Poisson resummation formula to evaluate the sum over $j$ in Eq.~\eqref{VacuumEnergy3}, applying it separately to each of the two terms~\cite{Kirsten:2010zp}. The formula is written as
\begin{equation}
\sum_{j=1}^{\infty}e^{-a j^2}
=-\frac{1}{2} + 
\sqrt{\frac{\pi}{4a}}
\sum_{j=-\infty}^{\infty}
e^{-\frac{\pi^2 j^2}{a}},
\label{PRF}
\end{equation}
for $a>0$.

We first consider the Landau-like contribution, $E_{\rm L}(s)$. By making use of Eq.~\eqref{iden}, the summation over $n$ of the exponential factor $e^{-\alpha(2n+1)\tau^2}$ can be carried out analytically~\cite{Junior:2025thl}, leading to
\begin{equation}
   E_{\rm L}(s) = \frac{A\alpha}{8\pi\Gamma(s)} \int_{0}^{\infty} \frac{d\tau \, \tau^{2s-1}e^{-m^2 \tau^2}}{\sinh(\alpha\tau^2)} \left[-1 + 
\frac{L}{\tau\pi^{\frac{1}{2}}}
+ \frac{2L}{\tau\pi^{\frac{1}{2}}}
\sum_{j=1}^{\infty}
e^{-\frac{L^2 j^2}{\tau^2}}\right],
   \label{LLS}
\end{equation}
where we have used Eq.~\eqref{PRF} with $a=\pi^2\tau^2/L^2$. The first term inside the square brackets can be integrated analytically, yielding a contribution proportional to $\alpha^{-s+1}\zeta_H(s,\gamma)$, where $\gamma=(\alpha+m^2)/(2\alpha)$ is the second argument of the Hurwitz zeta function. Taking the limit $s\rightarrow -1/2$, one finds that this contribution grows with $\alpha$. Such a behavior is physically unacceptable, since the vacuum energy is expected to vanish in the limit of large values of this parameter. Indeed, in analogy with the Landau-level problem, increasing the parameter $\alpha$ (or equivalently the magnetic field strength $eB$) suppresses quantum vacuum effects, and therefore the vacuum energy should not grow without bound. Notice also that, in the limits $\alpha\rightarrow\infty$ or $m\rightarrow0$, one has $\gamma\rightarrow1/2$, so that the Hurwitz zeta function reduces to $\zeta_H(-1/2,1/2)$, which is independent of $\alpha$. Consequently, the entire contribution scales as $\alpha^{\frac{3}{2}}$ for large values of $\alpha$, reinforcing its unphysical character. Moreover, this term is independent of the plate separation $L$, indicating that it corresponds to a bulk vacuum contribution rather than a boundary-induced effect. Therefore, it must be subtracted in the renormalization procedure used to define the vacuum energy.

Regarding the second term inside the square brackets in Eq.~\eqref{LLS}, the corresponding integral can also be evaluated analytically, yielding a contribution proportional to $L\alpha^{-s+\frac{3}{2}}\Gamma(s-1/2)\zeta_H(s-1/2,\gamma)$.
In the physical limit \(s\rightarrow -1/2\), the Hurwitz zeta function remains finite~\cite{Junior:2025thl}, whereas the gamma function develops a pole. Moreover, this contribution grows with both \(L\) and \(\alpha^2\), which is physically unacceptable, since the vacuum energy is expected to vanish as \(L\to\infty\). Therefore, this divergent contribution must also be removed through the renormalization procedure adopted to define the vacuum energy.

Upon denoting by \(E^{\rm sub}(s)\) the contributions removed in the renormalization procedure, including either divergent terms, finite but physically unacceptable contributions, or both, the renormalized vacuum energy can be formally defined as
\begin{equation}
E^{\rm ren}=\lim_{s\rightarrow-\frac{1}{2}}\big(E(s) - E^{\rm sub}(s)\big).
   \label{renVE}
\end{equation}

In the present case, following the scheme above, the renormalized Landau-like contribution is given by the last term inside the square brackets in Eq.~\eqref{LLS}, after subtracting $E_{\rm L}^{\rm sub}(s)$, which is composed of the first two terms in Eq.~\eqref{LLS}. Thus, by making the change of variable $\tau\rightarrow L\tau$, we obtain
\begin{eqnarray}
   \frac{E_{\rm L}^{\rm ren} }{A}&=& -\frac{\alpha_0}{8\pi^2L^3}\sum_{j=1}^{\infty} \int_{0}^{\infty} \frac{d\tau \, \tau^{-3}e^{-m_0^2 \tau^2 - \frac{j^2}{\tau^2}}}{\sinh(\alpha_0\tau^2)}\nonumber\\
   &=&-\frac{1}{8\pi^2L^3}\sum_{j=1}^{\infty}\mathcal{I}_L(j, m_0, \alpha_0),
   \label{LLS_ren}
\end{eqnarray}
where we have used $\Gamma(-1/2)=-2\sqrt{\pi}$ and defined the dimensionless parameters
\begin{equation}
   \alpha_0=\alpha L^2,\qquad\qquad m_0=mL.
   \label{alpha_mass}
\end{equation}
Note that the entire dependence on the parameter $\alpha$ is encoded in the function $\mathcal{I}_L(j,m_0,\alpha_0)$, which will be particularly useful in the asymptotic analysis presented later.

We next turn to the contribution, $E_{\rm c}(s)$, given by the second term in Eq.~\eqref{VacuumEnergy3}. Again, by making use of Eq.~\eqref{iden}, the summations over $n$ and $l$ of the exponential factor can be performed analytically, yielding
\begin{equation}
   E_{\rm c}(s) = \frac{A\alpha}{4\pi\Gamma(s)} \int_{0}^{\infty} \frac{d\tau \, \tau^{2s-1}e^{-m^2 \tau^2}}{\sinh(\alpha\tau^2)\big(e^{\alpha\tau^2} - 1\big)} \left[-1 + 
\frac{L}{\tau\pi^{\frac{1}{2}}}
+ \frac{2L}{\tau\pi^{\frac{1}{2}}}
\sum_{j=1}^{\infty}
e^{-\frac{L^2 j^2}{\tau^2}}\right],
   \label{AC}
\end{equation}
where we have used Eq.~\eqref{PRF}. The integrals associated with the first two contributions in Eq.~\eqref{AC} can, in principle, be evaluated analytically by leaving the sum over \(l\) in Eq.~\eqref{VacuumEnergy3} unevaluated after applying the identity~\eqref{iden}. As in the calculation leading to Eq.~\eqref{LLS}, the resulting expressions are written in terms of the gamma function and the Hurwitz zeta function \(\zeta_H(s,\gamma_l)\), with \(\gamma_l=\big(\alpha(1+l)+m^2\big)/2\alpha\). However, the subsequent sum over \(l\) of the Hurwitz zeta function is divergent. This divergence has a clear physical origin. The first integral in Eq.~\eqref{AC} corresponds to a bulk contribution, since it is independent of the plate separation \(L\), whereas the second grows linearly with \(L\). Neither term represents the interaction energy induced by the boundaries. In particular, the linear dependence on \(L\) is incompatible with the physical requirement that the boundary-induced vacuum energy vanish in the limit \(L\to\infty\). Therefore, both contributions must be removed through the renormalization procedure adopted to define the physical vacuum energy, leaving only the finite boundary-induced contribution.

The renormalized contribution in Eq.~\eqref{AC} arises from the third term inside the square brackets. Hence, after subtracting $E_{\rm c}^{\rm sub}(s)$, which is composed of the first two terms in Eq.~\eqref{AC}, we obtain
\begin{eqnarray}
   \frac{E_{\rm c}^{\rm ren}}{A} &=& -\frac{\alpha_0}{4\pi^2L^3}\sum_{j=1}^{\infty} \int_{0}^{\infty} \frac{d\tau \, \tau^{-3}e^{-m_0^2 \tau^2 - \frac{j^2}{\tau^2}}}{\sinh(\alpha\tau^2)\big(e^{\alpha\tau^2} - 1\big)} 
\nonumber\\
&=&-\frac{1}{8\pi^2L^3}\sum_{j=1}^{\infty} \mathcal{I}_{\rm c}(j, m_0, \alpha_0),
   \label{AC_ren}
\end{eqnarray}
where, as before, we have made the change of variable $\tau\rightarrow L\tau$.

Therefore, the total renormalized vacuum energy per unit area of the plates is given by
\begin{eqnarray}
 \frac{E_0^{\rm ren}}{A} =&-&\frac{1}{8\pi^2L^3}\sum_{j=1}^{\infty}\Big(\mathcal{I}_L(j, m_0, \alpha_0) + \mathcal{I}_{\rm c}(j, m_0, \alpha_0)\Big).
   \label{total_ren}
\end{eqnarray}
As we shall see next, the expression above is exponentially suppressed for large values of $\alpha_0$, whereas for small values of this parameter its leading contribution diverges, reflecting the restriction imposed by the quantization condition in Eq.~\eqref{integer_energy}. The next-to-leading term, however, exactly reproduces the vacuum energy associated with two perfectly reflecting parallel plates~\cite{Cougo-Pinto:1998jun, Cruz:2020zkc}.

We can consider now the asymptotic limits of the vacuum energy in Eq.~\eqref{total_ren} for $\alpha_0\ll 1$ and $\alpha_0\gg 1$. Starting by the latter, the dominant contribution for the functions $\mathcal{I}(j, m_0, \alpha_0)$ can be written as
\begin{eqnarray}
  \mathcal{I}(j, m_0, \alpha_0)&\simeq&2f\alpha_0\int_{0}^{\infty}d\tau \, \tau^{-3}e^{-m_0^2 \tau^2- \frac{j^2}{\tau^2} - f\alpha_0\tau^2}\nonumber\\
  &\simeq&\frac{2f\alpha_0\sqrt{f\alpha_0 + m_0^2}}{j}K_{1}\big(2j\sqrt{f\alpha_0 + m_0^2}\big)\nonumber\\
  &\simeq&\sqrt{\frac{2\pi f^3\alpha_0^3}{j^2}}e^{-2jf\alpha_0},
   \label{alphaB}
\end{eqnarray}
where in the last line we have considered $\alpha_0\gg m_0$ and used the asymptotic expression for large arguments $K_{a}(x)=\sqrt{\frac{\pi}{2}}e^{-x}$ for the Macdonald funtion~\cite{abramowitz1965handbook}. Note that $f=1$ for the function $\mathcal{I}(j, m_0, \alpha_0)$ in the first term on the r.h.s. of Eq.~\eqref{total_ren} while $f=2$ for the function in the second term. Note also that as $j$ is an index summation its main contribution to~\eqref{alphaB} is $j=1$. So, our conclusion here is that for very large values of the dimensionless parameter $\alpha_0$ the vacuum energy is exponentially suppressed for both terms in Eq.~\eqref{total_ren}. The term for $f=1$, however, dominates. This behaviour for the vacuum energy is expected since it must not grow without bound with a classical parameter like $\alpha$.

In contrast, for small values of $\alpha_0$, for both terms in Eq.~\eqref{total_ren} the function $\mathcal{I}(j, m_0, \alpha_0)$ can be generically write as
\begin{eqnarray}
  \mathcal{I}(j, m_0, \alpha_0)=f\int_{0}^{\infty}d\tau\tau^{-5} \, \frac{(\alpha_0\tau^2)}{g(\tau)}e^{-m_0^2 \tau^2- \frac{j^2}{\tau^2}},
   \label{alphaS}
\end{eqnarray}
where the function \(g(\tau)\) is defined by the denominators appearing in Eqs.~\eqref{LLS_ren} and~\eqref{AC_ren}, respectively. In a unified notation, it can be written as
\begin{equation}
g(w)=
\begin{cases}
\sinh\!(w)\simeq w + \frac{w^3}{6} + \mathcal{O}(w^5), \\[2mm]
\sinh\!(w)
\left(e^{w}-1\right)\simeq w^2 + \frac{w^3}{2} + \mathcal{O}(w^4),
\end{cases}
\label{fun_g}
\end{equation}
where, for each case, we have also displayed the first two nonvanishing terms in the series expansion for small \(w\). Substituting these expansions into Eq.~\eqref{alphaS}, we find that the contribution associated with Eq.~\eqref{LLS_ren} reduces to
\begin{eqnarray}
  \mathcal{I}_L(j,m_0,\alpha_0)
  =
  \frac{m_0^2K_2(2jm_0)}{j^2},
  \label{alphaS_R}
\end{eqnarray}
which, upon substitution into Eq.~\eqref{LLS_ren}, exactly reproduces the well-known vacuum contribution for two perfectly reflecting parallel plates~\cite{Cougo-Pinto:1998jun,Cruz:2020zkc}.

In contrast, making use of Eq.~\eqref{fun_g}, the integral defining
\(\mathcal{I}_{\rm c}(j,m_0,\alpha_0)\) in Eq.~\eqref{AC_ren} yields
\begin{eqnarray}
  \mathcal{I}_{\rm c}(j,m_0,\alpha_0)
  =
  \frac{2}{\alpha_0}
  \frac{m_0^2K_2(2jm_0)}{j^2},
  \label{alphaS_C}
\end{eqnarray}
which diverges as \(\alpha_0\to0\), reflecting the singular nature of this limit. This divergence can be traced back to the restriction imposed by the quantization condition in Eq.~\eqref{integer_energy}. 
\begin{figure}[htbp]
    \centering
    \includegraphics[width=0.497\textwidth]{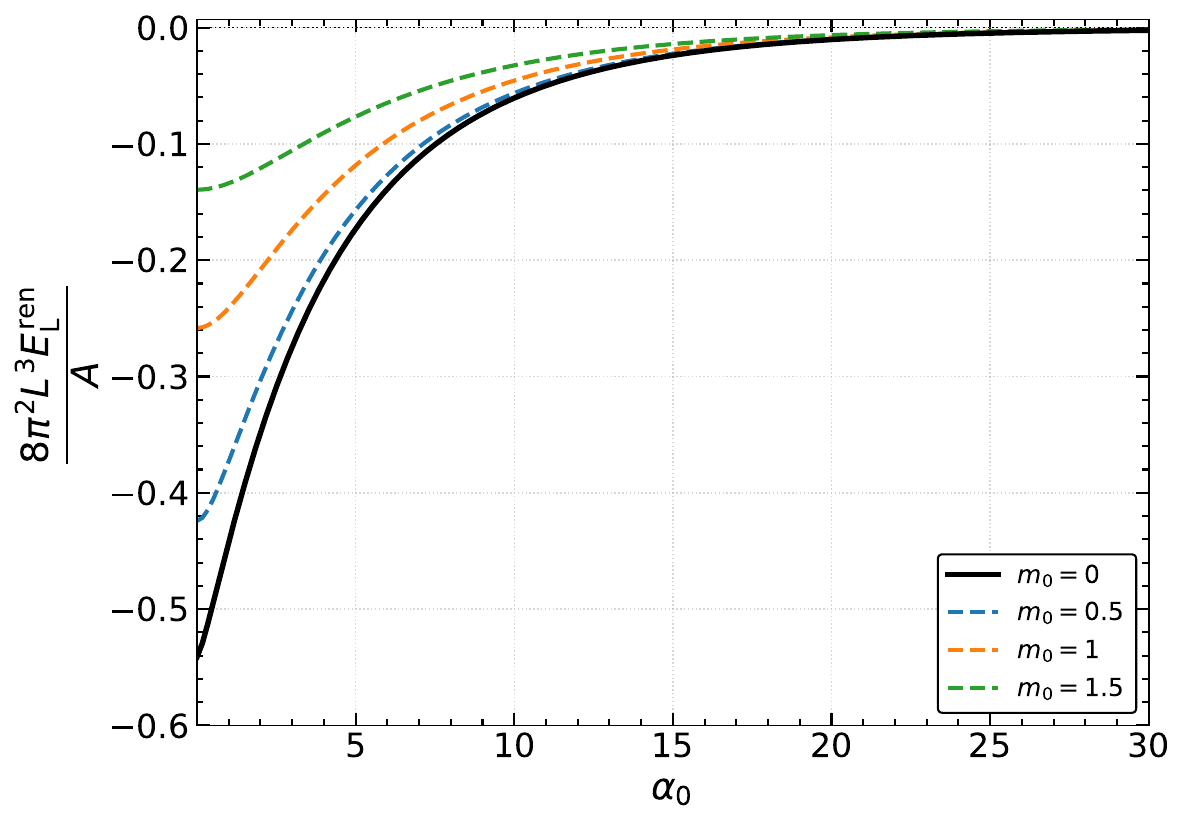} 
    \includegraphics[width=0.497\textwidth]{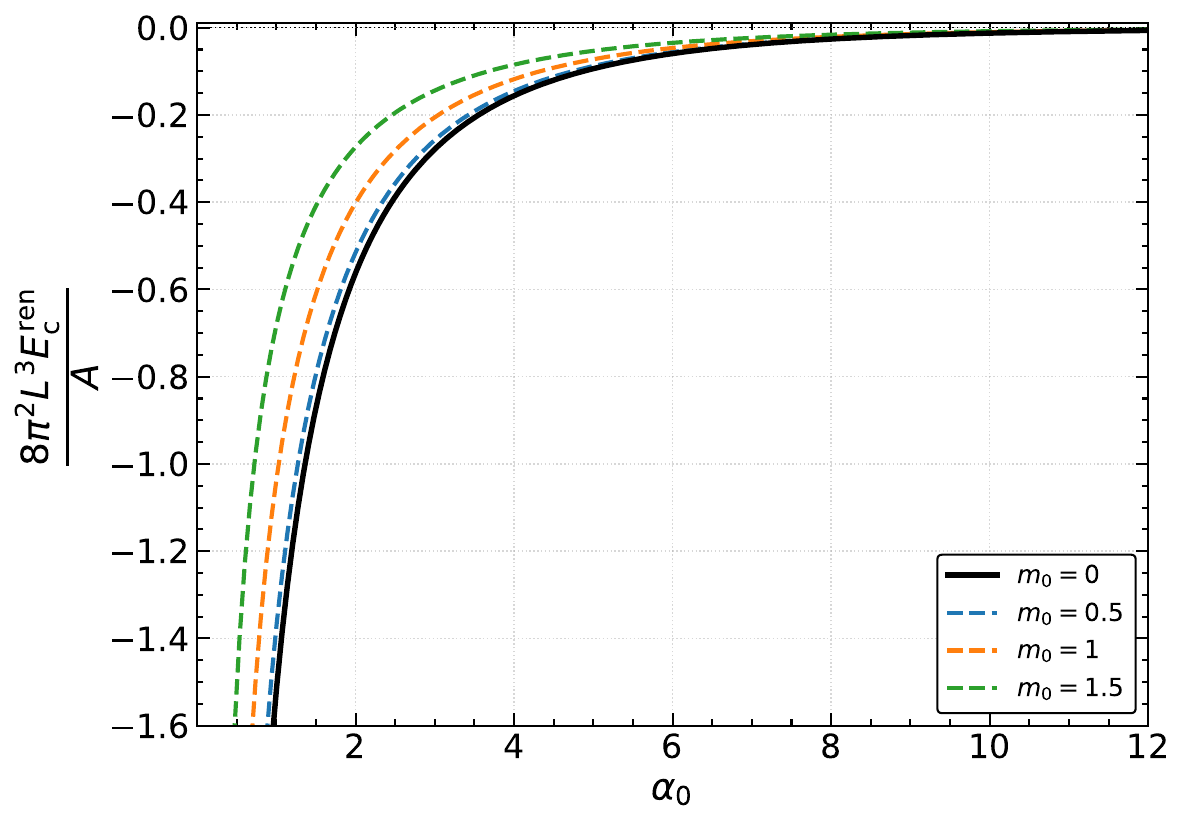} 
    \caption{Renormalized vacuum energy contributions as functions of the dimensionless parameter $\alpha_0$ for several values of the dimensionless mass parameter $m_0$. The left panel displays the contribution $E_{L}^{\rm ren}$, while the right panel displays the contribution $E_{\rm c}^{\rm ren}$. In both panels, the solid black curve corresponds to the massless case, $m_0=0$.}
    \label{fig:fig2}
\end{figure}

The numerical results for the two contributions in Eqs.~\eqref{LLS_ren} and~\eqref{AC_ren} to the renormalized vacuum energy are displayed in Fig.~\ref{fig:fig2}. The left panel shows the contribution $E_{\rm L}^{\rm ren}$, which has the same structure as the Landau-level contribution, differing only by an overall factor of two~\cite{Junior:2025thl}. This difference reflects the fact that the present model describes a real scalar field, whereas the conventional Landau-level spectrum corresponds to a complex scalar field, whose two independent degrees of freedom double the degeneracy of each level.

An important feature of $E_{\rm L}^{\rm ren}$ is that, in the limit $\alpha_0\to0$, it smoothly approaches the well-known vacuum energy for two perfectly reflecting parallel plates. This behavior is consistent with the analytical expansion discussed above and confirms that the present model correctly reproduces the standard boundary-induced vacuum energy in the absence of the position-dependent mass term.

The right panel displays the additional contribution $E_{\rm c}^{\rm ren}$, which has no counterpart in the Landau-level description. As anticipated from the asymptotic analysis, this contribution diverges as $\alpha_0\to0$. This divergence originates from the quantization condition~\eqref{integer_energy} adopted to obtain the exact spectrum, which is valid only for nonvanishing values of the parameter $\alpha$. Consequently, the limit $\alpha_0\to0$ cannot be taken directly in this contribution.

Finally, both panels clearly show that the renormalized vacuum energy is exponentially suppressed for increasing values of $\alpha_0$, as analytically demonstrated in Eq.~\eqref{alphaB}. This behavior indicates that the position-dependent mass progressively suppresses the vacuum fluctuations responsible for the Casimir interaction, rendering both contributions negligible in the regime of sufficiently large $\alpha_0$.

\begin{figure}[htbp]
    \centering
    \includegraphics[width=0.5\textwidth]{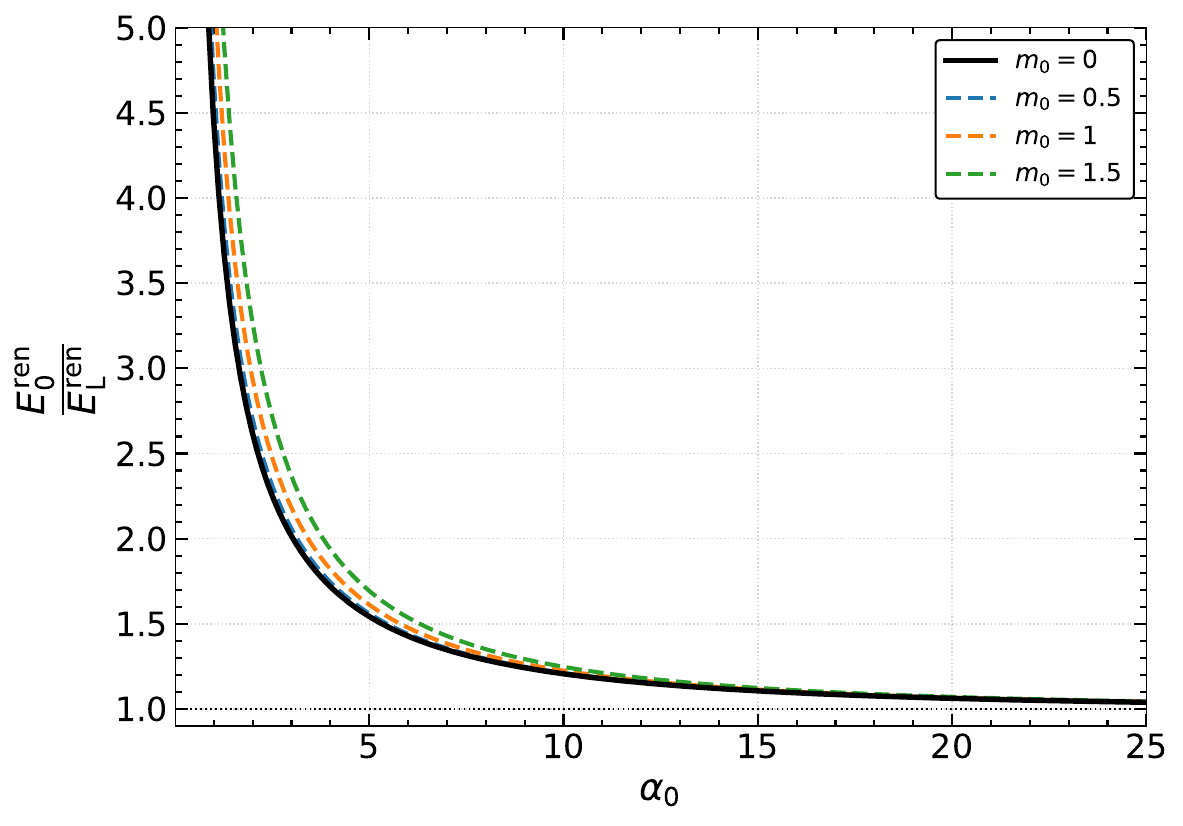} 
    \caption{Ratio between the total renormalized vacuum energy, $E_{0}^{\rm ren}$, and the contribution, $E_{\rm L}^{\rm ren}$, as a function of the dimensionless parameter $\alpha_0$ for several values of the dimensionless mass parameter $m_0$. The solid black curve corresponds to the massless case, $m_0=0$. The horizontal dotted line indicates the value unity, for which the additional contribution $E_{\rm c}^{\rm ren}$ vanishes.}
    \label{fig:fig3}
\end{figure}

Furthermore, Fig.~\ref{fig:fig3} displays the ratio between the total renormalized vacuum energy, $E_{0}^{\rm ren}$, and the Landau-level contribution, $E_{\rm L}^{\rm ren}$, both given by the expressions in Eqs.~\eqref{total_ren} and~\eqref{LLS_ren}. This quantity provides a direct measure of the relative importance of the additional contribution $E_{\rm c}^{\rm ren}$ introduced by the present model and given by Eq.~\eqref{AC_ren}.

As expected from the asymptotic analysis, the ratio increases significantly as $\alpha_0$ decreases. This behavior reflects the divergence of $E_{\rm c}^{\rm ren}$ in the limit $\alpha_0\rightarrow0$, which originates from the restriction imposed by Eq.~\eqref{integer_energy}. Consequently, in this regime the additional contribution cannot be regarded as a small correction to the Landau-level term.

On the other hand, for increasing values of $\alpha_0$ the ratio rapidly approaches unity for all values of the dimensionless mass parameter $m_0$. This demonstrates that the contribution $E_{\rm c}^{\rm ren}$ becomes progressively less relevant, so that the total renormalized vacuum energy is accurately described by the Landau-level contribution alone. Therefore, except in the vicinity of the singular limit $\alpha_0\rightarrow0$, the vacuum energy is dominated by the contribution $E_{\rm L}^{\rm ren}$.

\section{Conclusions}
\label{sec4}

In this work, we have investigated the Casimir effect for a massive real scalar field confined between two perfectly reflecting parallel plates in the presence of a quadratic position-dependent effective mass. The model provides a simple phenomenological realization of an external scalar background while preserving the local structure of the Klein-Gordon equation. At the relativistic level, the position-dependent mass discretizes the transverse spectrum, whereas in the nonrelativistic limit it naturally reproduces the dynamics of an harmonic oscillator.

After solving the Klein-Gordon equation exactly, we obtained the complete set of normal modes and showed that the corresponding energy spectrum possesses a characteristic Landau-like structure. Although the physical origin of the spectrum is entirely different from that of a charged particle in a uniform magnetic field, the parameter $\alpha$ plays a role analogous to the magnetic scale $eB$. Besides the usual Landau-type contribution, however, the present model gives rise to an additional term that has no counterpart in the conventional Landau problem and originates exclusively from the position dependence of the effective mass.

The quantization of the real scalar field allowed us to construct the vacuum energy associated with the confined modes. By employing generalized zeta-function regularization together with an appropriate renormalization prescription, we obtained finite expressions for the vacuum energy and identified the bulk and large-$L$ contributions that must be removed to isolate the physically meaningful boundary-induced energy.

The asymptotic analysis revealed two distinct physical regimes. For large values of the dimensionless parameter $\alpha_0=\alpha L^2$, both contributions to the renormalized vacuum energy become exponentially suppressed. This behavior reflects the progressive suppression of quantum vacuum fluctuations by the external scalar background and ensures that the vacuum energy does not increase without bound as the classical parameter $\alpha$ becomes large.

The opposite regime, $\alpha_0\ll1$, exhibits a richer structure. The Landau-like contribution smoothly reproduces the standard vacuum energy for a massive scalar field confined between parallel plates, demonstrating the consistency of the model with the known boundary-induced vacuum energy. In contrast, the additional contribution diverges in this limit. This divergence is not a failure of the renormalization procedure but rather a direct consequence of the quantization condition used to derive the spectrum, which is valid only for nonvanishing values of the parameter $\alpha$. Therefore, the limit $\alpha\rightarrow0$ is intrinsically singular for this sector of the spectrum.

The numerical analysis fully confirms the analytical predictions. In particular, both contributions exhibit exponential suppression for increasing values of $\alpha_0$, while the ratio between the total vacuum energy and the Landau-like contribution approaches unity. Consequently, except in the vicinity of the singular limit $\alpha\rightarrow0$, the vacuum energy is dominated by the Landau-like sector, with the additional contribution becoming progressively negligible.

The present work establishes a direct connection between position-dependent effective masses and vacuum phenomena induced by boundaries. Besides providing a new exactly solvable model for the Casimir effect, it opens the possibility of investigating more general spatially varying scalar backgrounds, different geometries and boundary conditions, finite-temperature effects, or analogous position-dependent mass models for fermionic and gauge fields.

\acknowledgments
We thank Knut Bakke for valuable discussions. R.L.A.X. and M.H.B.C. acknowledge financial support from the Brazilian National Council for Scientific and Technological Development (CNPq). E.R.B.M. and H.M. acknowledge partial financial support from CNPq under Grants No. 304332/2024-0 and No. 308049/2023-3, respectively.


\end{document}